\documentclass[10pt]{article}
\usepackage{epsfig}
\usepackage{axodraw}
\usepackage{amssymb}
\usepackage{graphicx}
\usepackage{wrapfig}
\usepackage{floatflt}
\begin{document}
\newcommand{\eqnzero}{\setcounter{equation}{0}} 

\newcommand{\bq}{\begin{equation}}
\newcommand{\eq}{\end{equation}}
\newcommand{\bqa}{\begin{eqnarray}}
\newcommand{\eqa}{\end{eqnarray}}
\newcommand{\baa}[1]{\begin{array}{#1}}
\newcommand{\eaa}{\end{array}}
\newcommand{\nll}{\nonumber\\}

\newcommand{\Litwo}{\mbox{${\rm{Li}}_{2}$}}
\newcommand{\alem}{\alpha_{em}}
\newcommand{\alsS}{\alpha^2_{_S}}
\newcommand{\ds }{\displaystyle}
\newcommand{\sss}[1]{\scriptscriptstyle{#1}}
\newcommand{\eps}{\varepsilon^*}
\newcommand{\dprop}{\overline\Delta}
\newcommand{\dpropi}[1]{d_{#1}}
\newcommand{\sla}[1]{/\!\!\!#1}
\def\mgn{mgn}
\def\mw {M_{\sss{W}}}
\def\mws{M_{\sss{W}^2}}
\def\mz {M_{\sss{Z}}}
\def\mh {M_{\sss{H}}}
\def\men{m_{\nu_e}}
\def\mel{m_e}
\def\mup{m_u}
\def\mdn{m_d}
\def\mmn{m_{\nu}}
\def\mmo{m_{\mu}}
\def\mch{mch}
\def\mst{mst}
\def\mtn{mtn}
\def\mta{mta}
\def\mtp{m_t}
\def\mbt{m_b}
\def\mp{mp}
\def\mf{m_f}
\def\mv{M_{\sss{V}}}
\def\srt{\sqrt{2}}
\newcommand{\vpa}[2]{\sigma_{#1}^{#2}}
\newcommand{\vma}[2]{\delta_{#1}^{#2}}
\newcommand{\af}{I^3_f}
\newcommand{\sqs}{\sqrt{s}}

\newcommand{\stw}{s_{\sss{W}}  }
\newcommand{\ctw}{c_{\sss{W}}  }
\newcommand{\stws}{s^2_{\sss{W}}}
\newcommand{\stwf}{s^4_{\sss{W}}}
\newcommand{\ctws}{c^2_{\sss{W}}}
\newcommand{\ctwf}{c^4_{\sss{W}}}

\newcommand{\siw }{s_{\sss{W}}}           
\newcommand{\cow }{c_{\sss{W}}}
\newcommand{\siws}{s^2_{\sss{W}}}
\newcommand{\cows}{c^2_{\sss{W}}}
\newcommand{\siwc}{s^3_{\sss{W}}}
\newcommand{\cowc}{c^3_{\sss{W}}}
\newcommand{\cowsc}{c^6_{\sss{W}}}
\newcommand{\siwf}{s^4_{\sss{W}}}
\newcommand{\cowf}{c^4_{\sss{W}}}

\newcommand{\bff}[4]{B_{#1}\big( #2;#3,#4\big)}             
\newcommand{\fbff}[4]{B^{F}_{#1}\big(#2;#3,#4\big)}        
\newcommand{\scff}[1]{C_{#1}}             
\newcommand{\sdff}[1]{D_{#1}}                 
\newcommand{\dffp}[6]{D_{0} \big( #1,#2,#3,#4,#5,#6;}       
\newcommand{\dffm}[4]{#1,#2,#3,#4 \big) }       
\newcommand{\tHmus}{\mu^2}
\newcommand{\epsh}{\hat\varepsilon}
\newcommand{\epsb}{\bar\varepsilon}


\newcommand{\chapt}[1]{Chapter~\ref{#1}}
\newcommand{\chaptsc}[2]{Chapter~\ref{#1} and \ref{#2}}
\newcommand{\eqn}[1]{Eq.~(\ref{#1})}
\newcommand{\eqns}[2]{Eqs.~(\ref{#1})--(\ref{#2})}
\newcommand{\eqnss}[1]{Eqs.~(\ref{#1})}
\newcommand{\eqnsc}[2]{Eqs.~(\ref{#1}) and (\ref{#2})}
\newcommand{\eqnst}[3]{Eqs.~(\ref{#1}), (\ref{#2}) and (\ref{#3})}
\newcommand{\eqnsf}[4]{Eqs.~(\ref{#1}), 
          (\ref{#2}), (\ref{#3}) and (\ref{#4})}
\newcommand{\eqnsv}[5]{Eqs.(\ref{#1}), 
          (\ref{#2}), (\ref{#3}), (\ref{#4}) and (\ref{#5})}
\newcommand{\tbn}[1]{Table~\ref{#1}}
\newcommand{\tabn}[1]{Tab.~\ref{#1}}
\newcommand{\tbns}[2]{Tabs.~\ref{#1}--\ref{#2}}
\newcommand{\tabns}[2]{Tabs.~\ref{#1}--\ref{#2}}
\newcommand{\tbnsc}[2]{Tabs.~\ref{#1} and \ref{#2}}
\newcommand{\fig}[1]{Fig.~\ref{#1}}
\newcommand{\figs}[2]{Figs.~\ref{#1}--\ref{#2}}
\newcommand{\figsc}[2]{Figs.~\ref{#1} and \ref{#2}}
\newcommand{\sect}[1]{Section~\ref{#1}}
\newcommand{\sects}[2]{Sections~\ref{#1} and \ref{#2}}
\newcommand{\subsect}[1]{Subsection~\ref{#1}}
\newcommand{\appendx}[1]{Appendix~\ref{#1}}

\begin{flushright}
{\tt hep-ph/0611188 \\ November 2006}
\end{flushright}
\vspace*{40mm}
\begin{center}
{\LARGE\bf Light-by-light scattering in {\tt SANC}}
\vspace*{15mm}

{\bf D.~Bardin$^{*}$, L.~Kalinovskaya$^{*}$, V.~Kolesnikov$^{*}$, E.~Uglov$^{*,**}$}
\vspace*{10mm}
{\normalsize{\it

$^{*}$  Dzhelepov Laboratory for Nuclear Problems, JINR,   \\
        ul. Joliot-Curie 6, RU-141980 Dubna, Russia;       \\
$^{**}$ The Faculty of Physics, Moscow State University,   \\
        Leninskie Gory, GSP-2, RU-119992, Moscow, Russia   }}
\vspace*{20mm}
\end{center}
\begin{abstract}
\noindent
In this paper we describe the implementation of the QED process $\gamma\gamma\to\gamma\gamma$ through a 
fermion loop into the framework of {\tt SANC} system. The computations of this process takes into account 
non-zero mass of loop-fermion. We briefly describe additional precomputation modules used for calculation
of massive fermion-box diagrams. We present the covariant and helicity amplitudes for this process and 
also some particular cases of $D_0$ and $C_0$ Passarino--Veltman functions. 
Whenever possible, we compare the results with those existing in the literature.
\end{abstract}
\vspace*{5mm}
\centerline{\it Talk presented at the International School--Seminar CALC2006, Dubna, 15-25 July 2006}
\vspace*{45mm}
\footnoterule
{\footnotesize \noindent
E-mails: bardin@nusun.jinr.ru, kalinov@nusun.jinr.ru,

\noindent
\hspace*{11mm} kolesnik@nusun.jinr.ru, corner@nusun.jinr.ru}
\clearpage

\section{Introduction \label{introduction}}
The {\tt SANC} is a computer system for semi-automatic calculations of realistic and pseudo-observables 
for various processes of elementary particle interactions "from the SM Lagrangian to event distributions" 
at the one-loop precision level for the present and future colliders --- TEVATRON, LHC, 
electron Linear Colliders (ISCLC, CLIC), muon factories and others. 
To learn more about available in {\tt SANC} processes see the description in~\cite{Andonov:2006} and
look at our home pages at JINR and CERN~\cite{SANC:2006}.

Light-by-light scattering is one of the most fundamental processes in QED. 
It proceeds via one-loop box diagrams containing charged particles. 
The first results for the low energy limit of this process were obtained by Euler~\cite{Euler:1936}.
Then Karplus and Neumann~\cite{Karplus:1951} found a solution for QED in general but complicated way.  
The QED cross sections in the high energy limit were calculated by Ahiezer~\cite{Ahiezer:1981}. 
Nowadays there are computations for $\gamma\gamma\to\gamma\gamma$ process in the electorweak Standard Model 
by Bohm \cite{Bohm:1994sf} and even for two-loop QCD and QED corrections by Bern~\cite{Bern:2001}.

In this paper we describe the implementation of the QED process $\gamma\gamma\to\gamma\gamma$ through 
fermion loop and corresponding precomputation block into the framework of {\tt SANC} system.
The computations of this process take into account non-zero mass of loop-fermions.

The paper is organized as follows:

First we discuss some notations and common expression for cross section in section~\ref{XS}. 

In section~\ref{CA} we discuss diagrams for $\gamma\gamma\to\gamma\gamma$ process and covariant amplitude 
tensor structure. 

In section~\ref{sbasis} one obtains compact form for this structure. 
The idea of form factors is described in section~\ref{FFs}.
 
The helicities amplitudes approach and their expressions for light-by-light scattering in general 
(massive) and in limiting (massless) cases are listed in section~\ref{HA}. 

In section \ref{precomputation} we shortly describe 
precomputation strategy \cite{Andonov:2006} and the place of this process on the {\tt SANC} tree. 

At last in section \ref{results} one can find the result in the limiting case and comparisons with those 
existing in the literature. 

Additionally, in Appendix section~\ref{appendix} we list answers for particular cases of $D_0$, $C_0$ 
and $B_0$ Passarino--Veltman (PV) functions~\cite{Passarino:1979} (see also the book~\cite{Bardin:1999}).
Finally, we present a table of integrals over the scattering angle, 
which are needed for calculation of light-by-light scattering through massive and massless loop fermions.

\section{Light-by-light scattering process \label{process}}
\subsection{Notation, cross section\label{XS}}
The 4-momenta of incoming photons are denoted by $k_1$ and $k_2$, of the outgoing ones 
--- by $k_3$ and $k_4$. 
The amplitudes are calculated for scattering of real photons, that is 
$k_1^2=0\,,~k_2^2=0\,,~k_3^2=0\,,~k_4^2=0$. The 4-momentum conservation law reads: 
\bqa
k_1+k_2-k_3-k_4=0\,.
\eqa
The Mandelstam variables are:\footnote{Note, that in {\tt SANC} we use Pauli metric.}
\bqa
&&s=-(k_1+k_2)^2=-2 k_1 \cdot k_2\,,\qquad t=-(k_1-k_3)^2=2 k_1 \cdot k_3\,,
\nll
&&u=-(k_1-k_4)^2=2 k_1  \cdot k_4\,,\qquad\qquad s+t+u=0\,.
\eqa
For the $2\to 2$ $\gamma\gamma\to\gamma\gamma$ process, the cross section has the form:
\bqa
d\sigma_{\gamma\gamma\to\gamma\gamma}=
\frac{1}{j}\left|{\cal{A}}_{\gamma\gamma\to\gamma\gamma}\right|^2 d\Phi^{(2)}\,,
\eqa
where $j=4\sqrt{(k_1k_2)^2}\,,$ is the flux,
${\cal{A}}_{\gamma\gamma\to\gamma\gamma}$ is the covariant amplitude (CA) of the process, 
and $d\Phi^{(2)}$ is the two body phase space:
\bqa
d\Phi^{(2)}=(2\pi)^4\delta\left(k_1+k_2-k_3-k_4\right) 
\frac{d^4 k_3 \delta\left( k_3^2 \right)}{(2\pi)^3}\frac{d^4k_4\delta\left( k_4^2 \right)}{(2\pi)^3}\,.
\eqa
For the differential cross section one gets: 
\bqa
d\sigma_{\gamma\gamma\to\gamma\gamma}=
\frac{1}{128\pi\omega^2}\left|{\cal{A}}_{\gamma\gamma\to\gamma\gamma}\right|^2 d \cos\theta\,,
\eqa
where
$\omega$ is the photons energy and $\theta$ --- the scattering angle in the center of mass system (CMS). 
\subsection{Covariant amplitude \label{CA}}
The covariant one-loop amplitude corresponds to a result of the straightforward standard calculation 
of all diagrams contributing to a given process at the tree (Born) and one-loop (1-loop) levels. 
\begin{small}
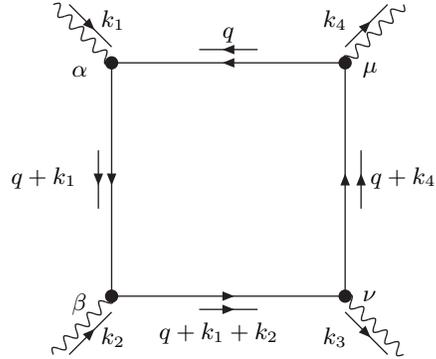
\begin{floatingfigure}{80mm}
\begin{picture}(132,132)(-60,+5)
 \Text(42,118)[lb]{$q$}    
 \Text(18,4)[lb]{$q+k_1+k_2$}
 \Text(-15,105)[lb]{$\alpha$}
 \Text(95,105)[lb]{$\mu$}
 \Text(95,20)[lb]{$\nu$}
 \Text(-15,15)[lb]{$\beta $}
 \Text(-4,123)[lb]{$k_1$}
 \Text(-4,3)[lb]{$k_2$}
 \Text(80,123)[lb]{$k_4$}
 \Text(80,3 )[lb]{$k_3$}
 \Text(98,63)[lb]{$q+k_4$}
 \Text(-38,63)[lb]{$q+k_1$}
 \ArrowLine(55,115)(33,115)
 \ArrowLine(33,17)(55,17)
 \ArrowLine(88,116)(104,132) 
 \ArrowLine(88,16)(104,0)
 \ArrowLine(94,55)(94,77)
 \ArrowLine(-16,0)(0,16)
 \ArrowLine(-16,132)(0,116) 
 \ArrowLine(-5,77)(-5,55)
 \Vertex(88,22){2.5}
 \ArrowLine(88,22)(88,110)
 \Vertex(88,110){2.5}
 \ArrowLine(88,110)(0,110)
 \Vertex(0,110){2.5}
 \ArrowLine(0,110)(0,22)
 \Vertex(0,22){2.5}
 \ArrowLine(0,22)(88,22)
 \Photon(88,22)(110,0){2}{5}
 \Photon(0,110)(-22,132){2}{5}
 \Photon(-22,0)(0,22){2}{5}
 \Photon(110,132)(88,110){2}{5}
\end{picture}
\caption[st-channel diagram for $\gamma\gamma\to\gamma\gamma$ process]
        {st-channel diagram for $\gamma\gamma\to\gamma\gamma$ process}\label{diagram.st}
\end{floatingfigure}
\end{small}
The CA is being represented in a certain basis, made of strings of Dirac matrices and/or 
external momenta (structures), contracted with polarization vectors of vector bosons, $\epsilon(k)$,
if any. 
The amplitude also contains kinematical factors and coupling constants and is parameterized by a certain 
number of Form Factors (FFs), which are denoted by ${\cal F}_{i}$, in general with an index $i$ labeling 
the corresponding structure. 
The number of FFs is equal to the number of independent structures.


The $\gamma\gamma\to\gamma\gamma$ process in QED appears due to non-linear effects of interaction with 
vacuum, so this process has no Born or tree level. Corresponding diagrams start from the one-loop level 
and in QED there are box diagrams with four internal fermions of equal mass. The number of not identical 
diagrams (or topologies) is equal to six. But three of them differ from another only by the orientation 
of the internal fermionic loop, giving the same contribution or a factor 2 to the amplitude. 
So, only three topologies (st, su and ut channels) remain which are related by simple permutations 
of external photons in the diagram shown in Figure~\ref{diagram.st}:
st-channel is shown, su-channel is obtained by $k_3 \leftrightarrow k_4$ and ut-channel --- by
$k_2 \leftrightarrow k_3$.

The full CA of given process for off-shell photons ($k_i\epsilon_i\neq 0$) 
can be written as:
\bq
{\cal A}_{\gamma\gamma\to\gamma\gamma}
  =4e^4Q^4_f\sum\limits_{i=1}^{43}{\cal F}_{i}\left(s\,,t\,,u\right) T_{i}^{\alpha\beta\mu\nu}\,,
\eq
where $e$ is the electron charge, $Q_f$ is the fraction of charge of loop fermion in units of electron 
charge, $T_{i}^{\alpha\beta\mu\nu}$ are tensors defined with an aid of auxiliary strings $\tau_{j}$ from 
the following subsection and ${\cal F}_{i}$ are FFs, depended on invariants $s\,,t\,,u$ and also on 
fermion mass and Passarino--Veltman functions. The off-shell process contains 43 basis elements.

\subsection{Strings and basis\label{sbasis}}
To obtain a compact form of structures of the amplitude we choose 14 auxiliary tensorial strings:

\def\lk{\hspace{-3mm}}
\bq
\begin{array}{llllllllll}
&\tau_{1}^{\alpha\beta} &\lk=\lk& k_{1\beta}k_{2\alpha}+\frac{1}{2}s\delta_{\alpha\beta}\,,
&\tau_{2}^{\mu\nu}      &\lk=\lk& k_{3\mu}k_{4\nu}+\frac{1}{2}s\delta_{\mu\nu}\,, 
&\tau_{3}^{\beta\nu}    &\lk=\lk& k_{2\nu}k_{3\beta}+\frac{1}{2}t\delta_{\beta\nu}\,,
\nll
&\tau_{4}^{\alpha\mu}   &\lk=\lk& k_{1\mu}k_{4\alpha}+\frac{1}{2}t\delta_{\alpha\mu}\,,
&\tau_{5}^{\alpha\nu}   &\lk=\lk& k_{1\nu}k_{3\alpha}+\frac{1}{2}u\delta_{\alpha\nu}\,,
&\tau_{6}^{\beta\mu}    &\lk=\lk& k_{4\beta}k_{2\mu}+\frac{1}{2}u\delta_{\beta\mu}\,,
\nll
&\tau_{7}^{\mu}         &\lk=\lk& k_{1\mu}-tu^{-1}k_{2\mu}\,,
&\tau_{8}^{\nu}         &\lk=\lk& k_{1\nu}-ut^{-1}k_{2\nu}\,,
&\tau_{9}^{\beta}       &\lk=\lk& k_{1\beta}-st^{-1}k_{3\beta}\,,
\nll
&\tau_{10}^{\alpha}     &\lk=\lk& k_{2\alpha}-su^{-1}k_{3\alpha}\,,
&\tau_{11}^{\mu}        &\lk=\lk& k_{4\mu}\,,~~\tau_{12}^{\nu}=k_{3\nu},
&\tau_{13}^{\beta}      &\lk=\lk& k_{2\beta}\,,~~\tau_{14}^{\alpha}=k_{1\alpha}.
\end{array}
\eq
The complete basis $T_{i}^{\alpha\beta\mu\nu}$ 
can be presented in a compact form with an aid of the auxiliary strings $\tau_{j}$:

\bq
\begin{array}{lllllllllllll}
&T^{\alpha\beta\mu\nu}_{1}  &\lk=\lk& \tau_{1}^{\alpha\beta}\tau_{2}^{\mu\nu},              
&T^{\alpha\beta\mu\nu}_{2}  &\lk=\lk& \tau_{3}^{\beta\nu}   \tau_{4}^{\alpha\mu},	      
&T^{\alpha\beta\mu\nu}_{3}  &\lk=\lk& \tau_{5}^{\alpha\nu}  \tau_{6}^{\beta\mu}, 	      
&T^{\alpha\beta\mu\nu}_{4}  &\lk=\lk& \tau_{1}^{\alpha\beta}\tau_{7}^{\mu}    \tau_{8}^{\nu},
\nll
&T^{\alpha\beta\mu\nu}_{5}  &\lk=\lk& \tau_{2}^{\mu\nu}     \tau_{9}^{\beta}  \tau_{10}^{\alpha}, 
&T^{\alpha\beta\mu\nu}_{6}  &\lk=\lk& \tau_{3}^{\beta\nu}   \tau_{7}^{\mu}    \tau_{10}^{\alpha},
&T^{\alpha\beta\mu\nu}_{7}  &\lk=\lk& \tau_{4}^{\alpha\mu}  \tau_{8}^{\nu}    \tau_{9}^{\beta},  
&T^{\alpha\beta\mu\nu}_{8}  &\lk=\lk& \tau_{5}^{\alpha\nu}  \tau_{7}^{\mu}    \tau_{9}^{\beta},  
\nll
&T^{\alpha\beta\mu\nu}_{9}  &\lk=\lk& \tau_{6}^{\beta\mu}   \tau_{8}^{\nu}    \tau_{10}^{\alpha},
&T^{\alpha\beta\mu\nu}_{28} &\lk=\lk& \tau_{6}^{\beta\mu}   \tau_{12}^{\nu}   \tau_{14}^{\alpha},
&T^{\alpha\beta\mu\nu}_{11} &\lk=\lk& \tau_{1}^{\alpha\beta}\tau_{7}^{\mu}    \tau_{12}^{\nu},   
&T^{\alpha\beta\mu\nu}_{12} &\lk=\lk& \tau_{2}^{\mu\nu}     \tau_{9}^{\beta}  \tau_{14}^{\alpha},
\nll
&T^{\alpha\beta\mu\nu}_{13} &\lk=\lk& \tau_{3}^{\beta\nu}   \tau_{7}^{\mu}    \tau_{14}^{\alpha}, 
&T^{\alpha\beta\mu\nu}_{14} &\lk=\lk& \tau_{4}^{\alpha\mu}  \tau_{8}^{\nu}    \tau_{13}^{\beta},  
&T^{\alpha\beta\mu\nu}_{15} &\lk=\lk& \tau_{5}^{\alpha\nu}  \tau_{7}^{\mu}    \tau_{13}^{\beta},  
&T^{\alpha\beta\mu\nu}_{16} &\lk=\lk& \tau_{6}^{\beta\mu}   \tau_{8}^{\nu}    \tau_{14}^{\alpha}, 
\nll			       
&T^{\alpha\beta\mu\nu}_{17} &\lk=\lk& \tau_{1}^{\alpha\beta}\tau_{11}^{\mu}   \tau_{8}^{\nu},     
&T^{\alpha\beta\mu\nu}_{18} &\lk=\lk& \tau_{2}^{\mu\nu}     \tau_{13}^{\beta} \tau_{10}^{\alpha}, 
&T^{\alpha\beta\mu\nu}_{19} &\lk=\lk& \tau_{3}^{\beta\nu}   \tau_{11}^{\mu}   \tau_{10}^{\alpha}, 
&T^{\alpha\beta\mu\nu}_{20} &\lk=\lk& \tau_{4}^{\alpha\mu}  \tau_{12}^{\nu}   \tau_{9}^{\beta},   
\nll	
&T^{\alpha\beta\mu\nu}_{21} &\lk=\lk& \tau_{5}^{\alpha\nu}  \tau_{11}^{\mu}   \tau_{9}^{\beta},   
&T^{\alpha\beta\mu\nu}_{22} &\lk=\lk& \tau_{6}^{\beta\mu}   \tau_{12}^{\nu}   \tau_{10}^{\alpha}, 
&T^{\alpha\beta\mu\nu}_{23} &\lk=\lk& \tau_{1}^{\alpha\beta}\tau_{11}^{\mu}   \tau_{12}^{\nu},    
&T^{\alpha\beta\mu\nu}_{24} &\lk=\lk& \tau_{2}^{\mu\nu}     \tau_{13}^{\beta} \tau_{14}^{\alpha}, 
\nll	
&T^{\alpha\beta\mu\nu}_{25} &\lk=\lk& \tau_{3}^{\beta\nu}   \tau_{11}^{\mu}   \tau_{14}^{\alpha},                      
&T^{\alpha\beta\mu\nu}_{26} &\lk=\lk& \tau_{4}^{\alpha\mu}  \tau_{12}^{\nu}   \tau_{13}^{\beta},	              
&T^{\alpha\beta\mu\nu}_{27} &\lk=\lk& \tau_{5}^{\alpha\nu}  \tau_{11}^{\mu}   \tau_{13}^{\beta}, 	              
&T^{\alpha\beta\mu\nu}_{10} &\lk=\lk& \tau_{7}^{\mu}        \tau_{8}^{\nu}    \tau_{9}^{\beta}     \tau_{10}^{\alpha},
\nll			       
&T^{\alpha\beta\mu\nu}_{29} &\lk=\lk& \tau_{7}^{\mu}        \tau_{8}^{\nu}    \tau_{13}^{\beta}    \tau_{14}^{\alpha}, 
&T^{\alpha\beta\mu\nu}_{30} &\lk=\lk& \tau_{7}^{\mu}        \tau_{9}^{\beta}  \tau_{12}^{\nu}      \tau_{14}^{\alpha}, 
&T^{\alpha\beta\mu\nu}_{31} &\lk=\lk& \tau_{7}^{\mu}        \tau_{10}^{\alpha}\tau_{12}^{\nu}      \tau_{13}^{\beta},  
&T^{\alpha\beta\mu\nu}_{32} &\lk=\lk& \tau_{8}^{\nu}        \tau_{9}^{\beta}  \tau_{11}^{\mu}      \tau_{14}^{\alpha}, 
\nll	
&T^{\alpha\beta\mu\nu}_{33} &\lk=\lk& \tau_{8}^{\nu}        \tau_{10}^{\alpha}\tau_{13}^{\beta}  \tau_{11}^{\mu},    
&T^{\alpha\beta\mu\nu}_{34} &\lk=\lk& \tau_{9}^{\beta}      \tau_{10}^{\alpha}\tau_{11}^{\mu}      \tau_{12}^{\nu},    
&T^{\alpha\beta\mu\nu}_{35} &\lk=\lk& \tau_{7}^{\mu}        \tau_{8}^{\nu}    \tau_{9}^{\beta}     \tau_{14}^{\alpha}, 
&T^{\alpha\beta\mu\nu}_{36} &\lk=\lk& \tau_{7}^{\mu}        \tau_{8}^{\nu}    \tau_{10}^{\alpha}   \tau_{13}^{\beta},  
\nll
&T^{\alpha\beta\mu\nu}_{37} &\lk=\lk& \tau_{7}^{\mu}        \tau_{9}^{\beta}  \tau_{10}^{\alpha}   \tau_{12}^{\nu},     
&T^{\alpha\beta\mu\nu}_{38} &\lk=\lk& \tau_{8}^{\nu}        \tau_{9}^{\beta}  \tau_{10}^{\alpha}   \tau_{11}^{\mu},     
&T^{\alpha\beta\mu\nu}_{39} &\lk=\lk& \tau_{11}^{\mu}       \tau_{12}^{\nu}   \tau_{13}^{\beta}    \tau_{10}^{\alpha},  
&T^{\alpha\beta\mu\nu}_{40} &\lk=\lk& \tau_{11}^{\mu}       \tau_{12}^{\nu}   \tau_{14}^{\alpha}   \tau_{9}^{\beta},    
\nll			       
&T^{\alpha\beta\mu\nu}_{41} &\lk=\lk& \tau_{11}^{\mu}       \tau_{13}^{\beta} \tau_{14}^{\alpha}   \tau_{8}^{\nu},
&T^{\alpha\beta\mu\nu}_{42} &\lk=\lk& \tau_{12}^{\nu}       \tau_{13}^{\beta} \tau_{14}^{\alpha}   \tau_{7}^{\mu},
&T^{\alpha\beta\mu\nu}_{43} &\lk=\lk& \tau_{11}^{\mu}       \tau_{12}^{\nu}   \tau_{13}^{\beta}    \tau_{14}^{\alpha} .
&                           &\lk \lk&
\end{array}
\label{basis}
\eq

Thus, one gets a minimal number of tensor structures of the CA, which contains a lot of 
terms. It can be written in an explicit form with an aid of scalar FFs. All masses and other parameters 
dependences are included 
into these FFs, but tensor structure with Lorenz indices are given by basis~(\ref{basis}).
It is important to emphasize that each basis elements $T^{\alpha\beta\mu\nu}_{i},\,i=1\div43$ is four
transversal with respect to each external photon:
\bqa
T^{\alpha\beta\mu\nu}_{i}k_{\alpha}=
T^{\alpha\beta\mu\nu}_{i}k_{\beta}=
T^{\alpha\beta\mu\nu}_{i}k_{\mu}=
T^{\alpha\beta\mu\nu}_{i}k_{\nu}=0\,.
\label{WI}
\eqa

\subsection{Form Factors \label{FFs}}
FFs are scalar coefficients in front of basis structures of the CA --- projections 
of CA to complete basis expressions $T_{i}^{\alpha\beta\mu\nu}$. They are presented as some combinations 
of scalar Passarino--Veltman functions $A_0$, $B_0$, $C_0$, $D_0$~\cite{Passarino:1979}.
They do not contain UV poles.

The number of terms in FFs equals to thousands in the case of non-zero mass of the loop fermion, 
but this number reduces greatly for zero loop fermion mass. 
Full answer for FFs one can find in the computer system {\tt SANC} on servers~\cite{SANC:2006}. 
For massless loop fermion the FFs are rather compact:
\bqa
{\cal F}_1 &=&\phantom{+}\frac{4}{3}\left(\frac{3i\pi}{s^2}+\frac{i\pi}{st}+\frac{i\pi}{su}
+\frac{\pi^2 ut}{2s^4}-\frac{3\pi^2}{4s^2}+\frac{1}{s^2}\right)
+\frac{4}{3}\left(-\frac{ut}{s^4}+\frac{3}{2s^2}\right)                                           l_t l_u 
\nll &&
+\frac{4}{3}\left(-\frac{3i\pi}{2s^2}-\frac{4i\pi}{su} 
         -\frac{i\pi}{u^2}-\frac{t}{s^3}+\frac{1}{s^2}+\frac{1}{su}\right)                        l_t
 -\frac{2}{3} \left(-\frac{ut}{s^4}+\frac{3}{s^2}+\frac{4}{su}+\frac{1}{u^2}\right)               l^2_t
 \nll &&
+\frac{4}{3} \left(-\frac{3i\pi}{2s^2}-\frac{4i\pi}{st}
                                 -\frac{i\pi}{t^2}-\frac{u}{s^3}+\frac{1}{s^2}+\frac{1}{st}\right)l_u 
 -\frac{2}{3} \left(-\frac{ut}{s^4}+\frac{3}{s^2}+\frac{4}{st}+\frac{1}{t^2}\right)               l^2_u\,, 
\nll
{\cal F}_2 &=&
\phantom{+}\frac{4}{3}\left(\frac{i\pi}{st}+\frac{i\pi}{su}-\frac{i\pi u}{t^3}-\frac{2i\pi}{t^2}
                    -\frac{\pi^2}{2s^2}-\frac{2\pi^2}{st}-\frac{3\pi^2}{4t^2}+\frac{1}{t^2}\right)
  +\frac{4}{3} \left(\frac{1}{s^2}+\frac{4}{st}+\frac{3}{2t^2}\right)                             l_t l_u 
\nll &&
+\frac{4}{3}\left(\frac{4i\pi}{st}+\frac{4i\pi}{su}
                         -\frac{3i\pi}{2t^2}-\frac{i\pi}{u^2}+\frac{1}{su}-\frac{3}{t^2}\right)   l_t
 -\frac{2}{3} \left(\frac{1}{s^2}-\frac{4}{su}+\frac{3}{t^2}+\frac{1}{u^2}\right)                 l^2_t   
\nll  &&
+\frac{4}{3} \left(\frac{i\pi su}{t^4}
               -\frac{3i\pi}{2t^2}+\frac{1}{st}-\frac{u}{t^3}+\frac{1}{t^2}\right)                l_u
 -\frac{2}{3} \left(\frac{1}{s^2}+\frac{4}{st}+\frac{u^2}{t^4}+\frac{u}{t^3}+\frac{3}{t^2}\right) l^2_u\,, 
\nll
 {\cal F}_3 &=&
\phantom{+}\frac{4}{3}\left(\frac{i\pi}{st}+\frac{i\pi}{su}-\frac{i\pi t}{u^3}-\frac{2i\pi}{u^2}
                   -\frac{\pi^2}{2s^2}-\frac{2\pi^2}{su}-\frac{3\pi^2}{4u^2}+\frac{1}{u^2}\right)
 +\frac{4}{3} \left(\frac{1}{s^2}+\frac{4}{su}+\frac{3}{2u^2}\right)                              l_t l_u 
 \nll &&
+\frac{4}{3}\left(\frac{4i\pi}{st}+\frac{4i\pi}{su}-\frac{3i\pi}{2u^2}
                                               -\frac{i\pi}{t^2}+\frac{1}{st}-\frac{3}{u^2}\right)l_u
 -\frac{2}{3} \left(\frac{1}{s^2}-\frac{4}{st}+\frac{3}{u^2}+\frac{1}{t^2}\right)                 l^2_u 
\nll &&
+\frac{4}{3} \left(-\frac{i\pi t}{u^3}-\frac{i\pi t^2}{u^4}
                               -\frac{3i\pi}{2u^2}+\frac{1}{su}-\frac{t}{u^3}+\frac{1}{u^2}\right)l_t
 -\frac{2}{3}\left(\frac{1}{s^2}+\frac{4}{su}+\frac{t^2}{u^4}+\frac{t}{u^3}+\frac{3}{u^2}\right)  l^2_t\,,
\nll
{\cal F}_5 &=& -\frac{4}{3}\left(\frac{3i\pi}{s^2}+\frac{2i\pi}{st}+\frac{2i\pi}{su}
                             +\frac{\pi^2 t^2}{2s^4}+\frac{\pi^2 t}{2s^3}-\frac{1}{s^2}\right)
 +\frac{4}{3} \left(\frac{t^2}{s^4}+\frac{t}{s^3}\right)                                          l_t l_u 
\nll &&
+\frac{4}{3}\left(\frac{3i\pi}{2s^2}+\frac{7i\pi}{2su}+\frac{2i\pi}{u^2}
                                                  -\frac{t}{s^3}-\frac{2}{s^2}-\frac{2}{su}\right)l_t
-\frac{2}{3} \left(\frac{t^2}{s^4}+\frac{t}{s^3}-\frac{3}{2s^2}-\frac{7}{2su}-\frac{2}{u^2}\right)l^2_t 
 \nll &&
+\frac{4}{3}\left(\frac{3i\pi}{2s^2}+\frac{7i\pi}{2st}+\frac{2i\pi}{t^2}
                                                  +\frac{t}{s^3}-\frac{1}{s^2}-\frac{2}{st}\right) l_u
 -\frac{2}{3} \left(\frac{t^2}{s^4}+\frac{t}{s^3}-\frac{3}{2s^2}-\frac{7}{2st}-\frac{2}{t^2}\right)l^2_u\,,
\nll
{\cal F}_7 &=&-\frac{4}{3}\left(\frac{i\pi}{st}+\frac{i\pi}{su}+\frac{2i\pi}{u^2}
                              +\frac{\pi^2 t}{s^3}+\frac{3\pi^2}{4s^2}-\frac{1}{su}\right)
 +\frac{4}{3}\left(\frac{2t}{s^3}+\frac{3}{2s^2}\right)                                    l_tl_u
 \nll &&
-\frac{4}{3}\left(\frac{2i\pi t}{u^3}+\frac{3i\pi}{2u^2}+\frac{2}{s^2}+\frac{1}{su}+\frac{2}{u^2}\right)
                                                                                           l_t
 -\frac{4}{3}\left(\frac{t}{s^3}+\frac{3}{4s^2}+\frac{t}{u^3}+\frac{3}{4u^2}\right)        l^2_t
\nll &&
+\frac{4}{3}\left(\frac{i\pi}{t^2}+\frac{2}{s^2}-\frac{1}{st}\right)                       l_u
 -\frac{4}{3}  \left(\frac{t}{s^3}+\frac{3}{4s^2}-\frac{1}{2t^2}\right)                    l^2_u\,,
\nll
{\cal F}_9 &=& -\frac{4}{3}\left(\frac{2i\pi}{st}+\frac{2i\pi}{su}+\frac{i\pi}{u^2}
           -\frac{\pi^2 t}{s^3}+\frac{3\pi^2}{4s^2}+\frac{3\pi^2}{4su}+\frac{1}{su}\right)
 +\frac{4}{3}\left(-\frac{2t}{s^3}+\frac{3}{2s^2}+\frac{3}{2su}\right)                     l_tl_u
\nll &&
+\frac{4}{3}\left(-\frac{i\pi t}{u^3}+\frac{2}{s^2}+\frac{1}{su}-\frac{1}{u^2}\right)      l_t 
 -\frac{4}{3}\left(-\frac{t}{s^3}+\frac{3}{4s^2}+\frac{3}{4su}+\frac{t}{2u^3}\right)       l^2_t
\nll &&
-\frac{4}{3}\left(\frac{3i\pi}{2st}+\frac{3i\pi}{2su}-\frac{2i\pi}{t^2}
                                      +\frac{2}{s^2}+\frac{2}{st}+\frac{3}{su}\right)      l_u
 -\frac{4}{3}\left(-\frac{t}{s^3}+\frac{3}{4s^2}+\frac{3}{4st}+\frac{3}{2su}-\frac{1}{t^2}\right)l^2_u\,, 
\nll
{\cal F}_{10} &=& \phantom{+}4\left(\frac{i\pi}{s^2}+\frac{i\pi}{st}+\frac{i\pi}{su}
         -\frac{\pi^2 t^2}{2s^4}-\frac{\pi^2 t}{2s^3}-\frac{\pi^2}{4s^2}-\frac{2}{s^2}\right)
 +4\left(\frac{t^2}{s^4}+\frac{t}{s^3}+\frac{1}{2s^2}\right)                                  l_tl_u 
\nll &&
-4\left(\frac{i\pi}{2s^2}+\frac{i\pi}{su}+\frac{i\pi}{u^2}+\frac{t}{s^3}-\frac{1}{su}\right)  l_t
 -2\left(\frac{t^2}{s^4}+\frac{t}{s^3}+\frac{1}{s^2}+\frac{1}{su}+\frac{1}{u^2}\right)        l^2_t
\nll &&
-4 \left(\frac{i\pi}{2s^2}+\frac{i\pi}{st}+\frac{i\pi}{t^2}+\frac{u}{s^3}-\frac{1}{st}\right) l_u
 -2 \left(\frac{t^2}{s^4}+\frac{t}{s^3}+\frac{1}{s^2}+\frac{1}{st}+\frac{1}{t^2}\right)       l^2_u\,,
\eqa
where
\bqa
l_t=\ln\left(-\frac{t}{s}\right),\qquad l_u=\ln\left(-\frac{u}{s}\right).
\eqa
Also there are equations among six FFs (even for massive case):
\bqa
{\cal F}_4 ={\cal F}_5 \,,\qquad
{\cal F}_6 = \frac{u^2}{t^2}{\cal F}_7 \,, \qquad
{\cal F}_8 = {\cal F}_9\,.
\eqa
The other FFs (namely 11-43) are not zero, but they do not contribute, 
because of corresponding basis structures for on-mass-shell photons satisfy Ward 
identity: $k_i\epsilon_i(k_i)=0$. 


\subsection{Helicity amplitudes \label{HA}}
In {\tt SANC} we use helicity amplitudes approach.

In the expression for CA as one can see in subsection~\ref{CA} one has tensor structures and a 
set of scalar FFs.
To calculate an observable quantity, such as cross section, one needs to make amplitude square, 
calculate products of Dirac spinors and contract Lorenz indices with polarization vector. 
In the standard approach making amplitude square 
one gets squares for each diagram and their interferences. This leads to a lot of terms.

In the helicity amplitudes approach we also derive tensor structure and FFs. But the next step 
is a projection to helicity basis and as a result one gets a set of non-interfering amplitudes, since
all of them are characterized by different set of helicity quantum numbers.
In this approach we can distinguish calculations of Dirac spinors and contraction of Lorenz indices 
from calculations of FFs. We can do this before making square of amplitude.
So, proceeding in this way, we get a profit on calculations time (less amount of terms due to zero 
interference) and also more clear step-by-step control.

For the process $\gamma\gamma\to\gamma\gamma$ one gets:
\bqa
{\cal A}_{\gamma\gamma\to\gamma\gamma}&=&4e^4Q^4_f\sum_{\rm{spins}}{\cal H}_{\rm{spins}}\,,\nll
\left|{\cal A}_{\gamma\gamma\to\gamma\gamma}\right|^2&=&16e^8Q^8_f\sum_{\rm{spins}}
\left|{\cal H}_{\rm{spins}}\right|^2\,.
\eqa
Note, the total number of HAs for this process is equal to 16. This corresponds to different combinations 
of external particles spin projections. For $\gamma\gamma\to\gamma\gamma$ processes there are 4 photons 
with two polarizations $'+'$ and $'-'$, so the total number is $2\cdot2\cdot2\cdot2=16$. 
Helicity amplitudes are scalar expressions.  
They result from an application of the procedure {\tt TRACEHelicity.prc}: 
\bqa
{\cal H}_{++++}={\cal H}_{----}&=&
\frac{1}{4}\bigg[s^2{\cal F}_{\sss 1}+t^2{\cal F}_{\sss 2}+u^2{\cal F}_{\sss 3}
       +2s^2{\cal F}_{\sss 5}+2su{\cal F}_{\sss 7}-2su{\cal F}_{\sss 9}+s^2{\cal F}_{\sss 10}\bigg]\,,\nll
{\cal H}_{+++-}={\cal H}_{++-+}&=& {\cal H}_{+-++}={\cal H}_{-+++} ={\cal H}_{---+}={\cal H}_{--+-}= \nll
{\cal H}_{-+--}={\cal H}_{+---}&=& 
\frac{1}{4} \bigg[-s^2{\cal F}_{\sss 5}-su{\cal F}_{\sss 7}+su{\cal F}_{\sss 9}
                               -s^2{\cal F}_{\sss 10}\bigg]\,,\nll
{\cal H}_{+-+-} = {\cal H}_{-+-+} &=&
 \frac{1}{4} \bigg[ u^2{\cal F}_{\sss 3} - 2su{\cal F}_{\sss 9} + s^2{\cal F}_{\sss 10} \bigg]\,, \nll
{\cal H}_{+--+} = {\cal H}_{-++-} &=&
 \frac{1}{4} \bigg[ t^2{\cal F}_{\sss 2} + 2su{\cal F}_{\sss 7} + s^2{\cal F}_{\sss 10} \bigg]\,, \nll
{\cal H}_{++--}={\cal H}_{--++}&=&
 \frac{1}{4} s^2 \bigg[ {\cal F}_{\sss 1} + 2{\cal F}_{\sss 5} +  {\cal F}_{\sss 10} \bigg] \,.
\eqa

Therefore, at this stage we observe five independent HAs, while
in the case of zero loop fermion mass one gets only four independent HAs which are very compact:
\bqa
&&\lk\lk\lk
{\ds {\cal H}_{++--} = {\cal H}_{--++}=
        -1+\left(\frac{t-u}{s}\right)\left(l_u-l_t\right)
                         -\left(\frac{1}{2}-\frac{ut}{s^2}\right) \left( \left(l_u - l_t\right)^2 + \pi^2 \right),
}
\nll
&&\lk\lk\lk
{\ds {\cal H}_{+-+-} = {\cal H}_{-+-+} = -1-i\pi\bigg(\frac{t-s}{u}\bigg)\,
-\Bigg[\left(1+i\pi\right)\bigg(\frac{t-s}{u}\bigg)\,+2i\pi\bigg(\frac{t}{u}\bigg)^2\Bigg] l_t\,
-\bigg(\frac{1}{2}-\frac{st}{u^2}\bigg) l^2_t \,,
}
\nll
&&\lk\lk\lk
{\ds
{\cal H}_{+--+} = {\cal H}_{-++-} =
-1-i\pi\bigg(\frac{u-s}{t}\bigg)
-\Bigg[\left(1+i\pi\right)\bigg(\frac{u-s}{t}\bigg)+2i\pi\bigg(\frac{u}{t}\bigg)^2\Bigg] l_u
-\bigg(\frac{1}{2}-\frac{su}{t^2}\bigg) l^2_u \,.}
\eqa
All the others HAs are equal to $+1$.


The relations among helicity HAs are due to C, P, T-invariance.
Moreover, another relation is due to crossing symmetry:
\bqa
{\cal H}_{+--+}\left(s,t,u\right)={\cal H}_{+-+-}\left(s,u,t\right)\,,
\eqa
but this fact does not mean reducing of the number of independent HAs.
\section{Process $\gamma\gamma\to\gamma\gamma$ in the SANC tree\label{precomputation}}
For boxes the {\tt SANC} idea of precomputation becomes vitally important~\cite{Andonov:2006}.
Calculation of some boxes for some particular processes takes so much time that an external user should 
refrain from repeating precomputation. Furthermore, the richness of boxes requires a classification. 
Depending on the type of external lines (fermion or boson), we distinguish three large classes 
of boxes: $ffff$, $ffbb$ and $bbbb$.

\clearpage

In this section we briefly describe modules relevant for $\gamma\gamma\to\gamma\gamma$. 
The sum of contributions of fermionic loop boxes form a gauge-invariant and UV-finite subset,
which is a consequence of Ward Identity \eqn{WI}. 

\begin{floatingfigure}{70mm}
\includegraphics[width=60mm,height=105mm]{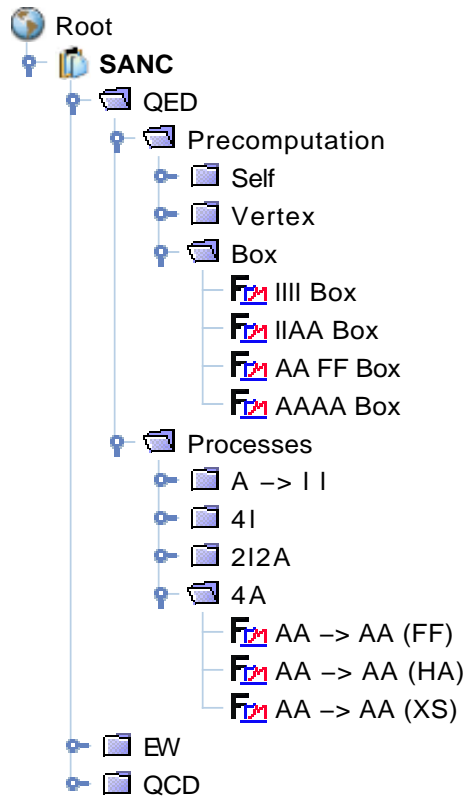}
\caption[$\gamma\gamma\to\gamma\gamma$ process in the QED tree]
        {$\gamma\gamma\to\gamma\gamma$ process in the QED tree\label{tree.QED}}
\end{floatingfigure}

The precomputation file {\tt AAAA Box} (see {\tt SANC} tree in Figure \ref{tree.QED}) contains
the sequence of procedures for calculation of the covariant amplitude. 
At this step we suppose, that all momenta are incoming (denoted by $p$'s) and photons are not 
on-mass-shell.
Therefore, these results can be used for other processes which need these parts as 
building blocks.

When we implement the process $\gamma\gamma\to\gamma\gamma$ (see {\tt 4A} branch), we use this 
building block several times by replacing incoming momenta $p$'s by corresponding kinematical
momenta $k$'s, 
and calculate FFs by the module {\tt AA->AA (FF)}, then helicity amplitudes by the module {\tt AA->AA (HA)}
and finally --- the differential and total process cross section by the module {\tt AA->AA (XS)}.

Also in {\tt SANC} system one has an opportunity of sending the analytical results to numerical 
evolution~\cite{Andonov:2006}.

\vspace*{35mm}


\section{Results and comparison\label{results}}
Now let us summarize the result.

The differential cross section of $\gamma\gamma\to\gamma\gamma$ process in QED has a form:
\bq
d\sigma_{\gamma\gamma\to\gamma\gamma} = 
\frac{e^8}{8\pi\omega^2} \sum_{\rm{spins}}\left|{\cal H}_{\rm{spins}}\right|^2 d\cos\theta\,,
\eq
where $\omega$ is the photons frequency, $\theta$ is the scattering angle in CMS and helicity amplitudes 
are expressed in terms of FFs. 
All dependences on Mandelstam invariants and loop fermion mass, also from Passarino-Veltman functions 
are included into these FFs.
All intermediate results for this process can be found in the {\tt SANC} system on the project 
servers~\cite{SANC:2006}.
The final answer for the total cross section in the massless limit after substitutions of helicity 
amplitudes and angular integration has a form:
\bq
\sigma_{\gamma\gamma\to\gamma\gamma} = \frac{e^8}{2\pi\omega^2} \left(\frac{108}{5} 
    + \frac{13}{2}\pi^2 - 8\pi^2\zeta(3) + \frac{148}{225}\pi^4 - 24\zeta(5) \right).
\eq
This result was compared with \cite{Ahiezer:1981} and the complete agreement was found.
Also the limit of helicity amplitudes was compared separately
with~\cite{Bern:2001} and again full agreement was observed.

One should emphasize that the obtained building block for box diagram in QED  
(as in $\gamma\gamma\to\gamma\gamma$) is the first step in the creation 
of environment for calculations of similar processes in the Standard Model 
(like $\gamma\gamma\to\gamma\gamma$, $\gamma\gamma\to ZZ$~\cite{Diakonidis:2006})
and in QCD (like $gg\to \gamma\gamma$, $gg\to ZZ$, $gg\to W^{+}W^{-}$ etc.)
\vspace*{5mm}

\section*{Acknowledgments \label{acknowledgements}}
\addcontentsline{toc}{section}{Acknowledgments}

\vspace*{5mm}
\noindent
The authors are grateful to G.~Nanava for useful discussions of helicity amplitudes 
and to A.~Arbuzov for providing us with useful references.

\section{Appendix \label{appendix}}
To obtain the cross section of the $\gamma\gamma\to\gamma\gamma$ process in an
analytic form we have to compute in an explicit form the master integrals, $B_0$, $C_0$, $D_0$ ---
scalar PV functions \cite{Passarino:1979}, \cite{Bardin:1999}  --- for a particular set of parameters. 
In $D_0$ and $C_0$ functions one can see collinear divergences, but 
the differential cross section is free of mass singularities which completely
cancel in the sum of all terms. 
The $A_0$ and $B_0$ functions contain UV divergences, which cancel in the sum of box contributions. 
In the process of computation we face also a problem of the ``angular edge'' divergences, but they are not 
physical and also cancel completely.

\subsection{$B_0$ function}
The $B_0$ function for $\gamma\gamma\to\gamma\gamma$ process reads:
\bqa
B_0\left(Q^2;M,M\right)&=&\frac{1}{\bar\epsilon}+2-\ln\left(\frac{M^2}{\mu^2}\right)
              -\beta\ln\left(\frac{\beta+1}{\beta-1}\right)\,,
\eqa
where $$\beta^2=1+\frac{4\widetilde{M}^2}{Q^2}\,,$$
and
\bqa
\widetilde{M}^2=M^2-{i}\epsilon\,.
\eqa
In zero limit of fermion mass: $M\to0$, one gets:
\bqa
B_0\left(-s;M,M\right)&=&\frac{1}{\bar\epsilon}+2-\left[\ln\left(\frac{s}{\mu^2}\right)-{i}\pi\right],
\nll
B_0\left(-t;M,M\right)&=&\frac{1}{\bar\epsilon}+2-\left[l_t+\ln\left(\frac{s}{\mu^2}\right)\right],
\nll
B_0\left(-u;M,M\right)&=&\frac{1}{\bar\epsilon}+2-\left[l_u+\ln\left(\frac{s}{\mu^2}\right)\right].
\eqa

\subsection{$C_0$ function}
The $C_0$ function is:
\bqa
C_0\left(0,0,Q^2;M,M,M\right)=
\int\limits_{0}^{1}dx\int\limits_{0}^{x}dy\left(Q^2y-Q^2xy+\widetilde{M}^2\right)^{-1},
\eqa

\noindent
After calculations:
\bqa
C_0\left(0,0,Q^2;M,M,M\right)=-\frac{1}{Q^2}\left[\Litwo\left(\frac{1}{x_1}\right)
                                                 +\Litwo\left(\frac{1}{x_2}\right)\right],
\eqa
where
\bqa
x_{1,2}=\frac{1}{2}\left(1\pm\beta\right).
\eqa

\noindent
For $M\to0$:
\bqa
C_0\left(0,0,-s;M,M,M\right)&=&-\frac{1}{2s}\left[\ln\left(\frac{M^2}{s}\right)+{i}\pi\right]^2,
\nll
C_0\left(0,0,-t;M,M,M\right)&=&-\frac{1}{2t}\left[\ln\left(\frac{M^2}{s}\right)-l_t\right]^2,
\nll
C_0\left(0,0,-u;M,M,M\right)&=&-\frac{1}{2u}\left[\ln\left(\frac{M^2}{s}\right)-l_u\right]^2.
\eqa

\subsection{$D_0$ function}
The $D_0$ function for st-channel diagram looks like:
\bqa
D_0\left(0,0,0,0,-s,-t;M,M,M,M \right)=
\int\limits_{0}^{1}dx \int\limits_{0}^{x}dy 
       \int\limits_{0}^{y}dz \left(txy-\left(s+t\right)xz+syz-ty+tz+\widetilde{M}^2\right)^{-2}
\eqa
After lengthy calculations:
\bqa
     && D_0\left(0,0,0,0,-s,-t;M,M,M,M\right) = -\frac{2}{stA_3}\Biggl\{
\nll &&
+2\pi{i}\int\limits_{0}^{1}dx\left[\frac{\eta\left(A_1+A_3\,,\ds\frac{x+A_1}{A_1+A_3}\right)}{x-A_3}
-\frac{\eta\left(A_1-A_3\,,\ds\frac{x+A_1}{A_1-A_3}\right)}{x+A_3}\right]
\nll &&
+2\pi{i}\int\limits_{0}^{1}dx\left[\frac{\eta\left(-A_1+A_3\,,\ds\frac{x-A_1}{-A_1+A_3}\right)}{x-A_3}
-\frac{\eta\left(-A_1-A_3\,,\ds\frac{x-A_1}{-A_1-A_3}\right)}{x+A_3}\right]
\nll &&
+2\pi{i}\int\limits_{0}^{1}dx\left[\frac{\eta\left(A_2+A_3\,,\ds\frac{x+A_2}{A_1+A_3}\right)}{x-A_3}
-\frac{\eta\left(A_2-A_3\,,\ds\frac{x+A_2}{A_2-A_3}\right)}{x+A_3}\right]
\nll &&
+2\pi{i}\int\limits_{0}^{1}dx\left[\frac{\eta\left(-A_2+A_3\,,\ds\frac{x-A_2}{-A_2+A_3}\right)}{x-A_3}
-\frac{\eta\left(-A_2-A_3\,,\ds\frac{x-A_2}{-A_2-A_3}\right)}{x+A_3}\right]
\nll &&
-\Biggl(\ln\left(1+A_3\right)-\ln\left(A_3\right)\Biggr)
   \Biggl[\ln\left(\frac{s}{4\widetilde{M}^2}\right)+\ln\left(\frac{t}{4\widetilde{M}^2}\right)
     +\ln\left(A_1-A_3\right)
\nll &&
 +\ln\left(-A_3-A_1\right)+\ln\left(A_2-A_3\right)+\ln\left(-A_3-A_2\right)\Biggr]
\nll &&
+\Biggl(\ln\left(-A_3+1\right)-\ln\left(-A_3\right)\Biggr)
   \Biggl[\ln\left(\frac{s}{4\widetilde{M}^2}\right)
+\ln\left(\frac{t}{4\widetilde{M}^2}\right)+\ln\left(A_1+A_3\right)
\nll &&
+\ln\left(A_3-A_1\right)+\ln\left(A_2+A_3\right)+\ln\left(A_3-A_2\right)\Biggr]
\nll &&
+\Litwo\left(-\frac{1+A_3}{-A_1-A_3}\right)-\Litwo\left(-\frac{1-A_3}{-A_1+A_3}\right) 
  +\Litwo\left(-\frac{1+A_3}{A_1-A_3}\right)-\Litwo\left(-\frac{1-A_3}{A_1+A_3}\right)
\nll &&
+\Litwo\left(-\frac{1+A_3}{-A_2-A_3}\right)-\Litwo\left(-\frac{1-A_3}{-A_2+A_3}\right) 
  +\Litwo\left(-\frac{1+A_3}{A_2-A_3}\right)-\Litwo\left(-\frac{1-A_3}{A_2+A_3}\right)\Biggr\},
\eqa
where
\bqa
 A_1=\sqrt{1-\frac{4\widetilde{M}^2}{s}}\,,~~
 A_2=\sqrt{1-\frac{4\widetilde{M}^2}{t}}\,,~~
 A_3=\sqrt{1-\frac{4\widetilde{M}^2\left(s+t\right)}{st}}.
\eqa
Veltman eta-function in terms of $\theta\left(x\right)$ functions of Heaviside has a form:
\bqa
\eta\left(A\,,B\right)=2\pi i \left(\theta\left(-ImA\right)\theta\left(-ImB\right)\theta\left(ImAB\right)-\theta\left(ImA\right)\theta\left(ImB\right)\theta\left(-ImAB\right)\right)\,.
\eqa

\noindent
For $M\to0$:
\bqa
D_0\left(0,0,0,0,-s,-t;M,M,M,M\right)=\frac{2}{st}\left[ \ln^2\left(-\frac{M^2}{t}\right)
 + \ln\left(-\frac{M^2}{t}\right) l_t -\frac{\pi^2}{2}+i\pi \ln\left(-\frac{M^2}{t}\right) \right].
\eqa
For two other channels we present only limiting case:
\bqa
\lk\lk
D_0\left(0,0,0,0,-u,-t;M,M,M,M\right) &\lk=\lk& \frac{2}{ut}\left[ \ln^2\left(\frac{M^2}{s}\right) 
 - \ln\left(\frac{M^2}{s}\right) \left(l_t+l_u\right) 
-\frac{\pi^2}{2}+l_t l_u \right],
\nll
\lk\lk
D_0\left(0,0,0,0,-s,-u;M,M,M,M\right) &\lk=\lk& \frac{2}{su}\left[ \ln^2\left(-\frac{M^2}{u}\right)
+ \ln\left(-\frac{M^2}{u}\right) l_u -\frac{\pi^2}{2}+i\pi \ln\left(-\frac{M^2}{u}\right) \right].
\eqa

\subsection{Table of integrals over scattering angle}
To calculate the total cross section one needs to integrate differential cross section over the angle 
$\theta$. 
We make substitution $x = \cos\theta$:
\bqa
 \frac{t}{s} = - \frac{1+x}{2}\,,\qquad
 \frac{u}{s} = - \frac{1-x}{2}\,,\quad
 -1<x<+1.
\eqa
The result for the total cross section is obtained with an aid of the table of integrals:
\bqa
&&\int\limits_{-1}^{+1}\ln\left(x\right)\ln^3\left(1-x\right)\left[-2+8x-16x^2+16x^3-8x^4\right]dx=
\nll && \hspace*{5cm}
 -\frac{1}{75}\left(\frac{229351664}{108000}-\frac{14\pi^4}{3}-\frac{18989\pi^2}{180}
-494\zeta\left(3\right)\right),\nll
&&\int\limits_{-1}^{+1}\ln^3\left(1-x\right)\left[\frac{8}{x^3}-\frac{12}{x^2}
+ \ln\left(1-x\right)\left(\frac{2}{x^4}-\frac{4}{x^3}+\frac{4}{x^2}\right)\right]dx
= -\frac{8\pi^2}{3}+\frac{32\pi^4}{45}+24\zeta\left(3\right),\nll
&&\int\limits_{-1}^{+1}\ln^2\left(x\right)\ln^2\left(1-x\right)\left[ \frac{1}{2}-2x+4x^2-4x^3+2x^4 
\right]dx=
\nll && \hspace*{5cm}
+\frac{1}{225}\left(-\frac{12239\pi^2}{180}
               +\frac{430069869}{324000}-494\zeta\left(3\right)-\frac{7\pi^4}{12} \right),\nll
&&\int\limits_{-1}^{+1}\ln\left(x\right)\ln^2\left(1-x\right)\left[ 6-24x+36x^2-24x^3 \right] dx=
-\frac{5\pi^2}{6} +\frac{1253}{144},\nll
&&\int\limits_{-1}^{+1}\ln^4\left(1-x\right)\left[ 1-2x+4x^2-4x^3+2x^4 \right] dx 
                                      = \frac{184815041}{8100000},\nll
&&\int\limits_{-1}^{+1}\ln^3\left(1-x\right)\left[ -4+8x-12x^2+8x^3 \right] dx= \frac{331}{144},\nll
&&\int\limits_{-1}^{+1}\ln\left(x\right)\ln\left(1-x\right)\left[4-12x+12x^2
+\pi^2\left(1-2x\right)^2+4\pi^2x^2\left(1-x\right)^2 \right] dx=
\frac{1}{9}\left(35+\frac{7739\pi^2}{1500}-\frac{7\pi^4}{10}\right),\nll
 &&\int\limits_{-1}^{+1}\ln^2\left(1-x\right)\left( 8-12x+12x^2+\pi^2\left(3-4x+8x^2-8x^3+4x^4\right) 
\right)dx=
\frac{1}{9}\left(125+ \frac{72989\pi^2}{1500}\right),\nll
 &&\int\limits_{-1}^{+1}\ln^2\left(1-x\right)\left[\frac{1}{x^4}-\frac{2}{x^3}\right]dx  
 + \int\limits_{-1}^{+1}\ln\left(1-x\right)\left[ \frac{2}{x^3}-\frac{3}{x^2} \right]dx 
 + \int\limits_{-1}^{+1}\left[\frac{1}{x^2} -\frac{1}{x}\right]dx=
\frac{1}{3}\left( 1 - \frac{2\pi^2}{3} \right),\nll
&&\int\limits_{-1}^{+1}\ln\left(1-x\right)\left[\frac{4}{x}-4+4x
                                                       +\pi^2\left(\frac{8}{x}-3+4x-6x^2+4x^3\right) 
\right]dx = 1+\frac{11\pi^2}{12}-\frac{4\pi^4}{3},\nll
&&\int\limits_{-1}^{+1}\left[\pi^2\left(3-2x+2x^2\right)+8+\pi^4\left(\frac{1}{4}-x+2x^2-2x^3+x^4\right) 
\right]dx= \frac{7\pi^4}{60}+\frac{8\pi^2}{3}+8,\nll
&&\int\limits_{-1}^{+1}\ln^4\left(1-x\right)\frac{1}{x}dx = 24\zeta\left(5\right),
~~~~~~~\int\limits_{-1}^{+1}\ln^3\left(1-x\right)\frac{1}{x}dx = -\frac{\pi^4}{15},\nll
 &&\int\limits_{-1}^{+1}\ln^2\left(1-x\right)\frac{1}{x^2}dx = \frac{\pi^2}{3},
~~~~~~~~~~~\int\limits_{-1}^{+1}\ln^2\left(1-x\right)\frac{1}{x}dx = 2\zeta\left(3\right).\nonumber
\eqa
\def\href#1#2{#2}
\addcontentsline{toc}{section}{References}
\end{document}